\documentclass[sigconf]{acmart}

\usepackage[nolist]{acronym}
\usepackage{bytefield}
\usepackage{tikz}
\usepackage{pgf-umlcd}
\usepackage{subfigure}
\usepackage{booktabs} 
\usepackage{flushend}

\newcommand{\etal}{\textit{et al}. }

\newcommand{\eg}{\textit{e}.\textit{g}. }

\setcopyright{usgov}

\acmDOI{10.1145/3199902.3199907}

\acmISBN{978-1-4503-6413-3/18/06}

\acmConference[WNS3 2018]{2018 Workshop on ns-3}{June 2018}{Surathkal, India}
\acmBooktitle{Proceedings of the 2018 Workshop on ns-3 (WNS3 2018), June 2018, Surathkal, India}
\acmYear{2018}
\copyrightyear{2018}

\acmPrice{15.00}

\thanks{Approved for public release: distribution unlimited.}

\begin{document}
\title{Implementation of Epidemic Routing with \\ IP Convergence Layer in ns-3}

\author{Justin P. Rohrer and Andrew N. Mauldin}
\orcid{1234-5678-9012}
\affiliation{%
  \institution{Naval Postgraduate School}
  \city{Monterey}
  \state{California}
  \postcode{93943-5285} \\
  \url{https://tancad.net/}
}
\email{jprohrer@nps.edu}


\begin{acronym}[XXXXXXX] 
	\acro{AEB}{Adaptive Exponential Beacon Protocol}
	\acro{AODV}{Ad-hoc On-demand Distance Vector}
	\acro{ASCII}{American Standard Code for Information Interchange}
	\acro{BGP}{Border Gateway Protocol}
	\acro{CAR}{Context Aware Routing Protocol}
	\acro{DECA}{Density Aware Reliable Broadcasting Protocol}
	\acro{DOD}{Department of Defense}
	\acro{DTN}{Disruption Tolerant Network}
	\acro{DTNRG}{Delay Tolerant Networking Research Group}
	\acro{DSDV}{Destination Sequence Distance Vector}
	\acro{DSR}{Dynamic Source Routing}
	\acro{EIGRP}{Enhanced Interior Gateway Routing Protocol}
	\acro{FIFO}{First In First Out}
	\acro{FTP}{File Transfer Protocol}
	\acro{GAPR}{Geolocation Aware Routing Protocol}
	\acro{GAPR2}{Geolocation Aware Routing Protocol 2}
	\acro{GAPR2a}{Geolocation Aware Routing Protocol 2a}
	\acro{GPS}{Global Positioning System}
	\acro{GUI}{Graphical User Interface}
	\acro{ICMP}{Internet Control Message Protocol}
	\acro{IP}{Internet Protocol}
	\acro{IRTF}{Internet Research Task Force}
	\acro{JiST}{Java in Simulation Time}
	\acro{LCAC}{Landing Craft Air Cushion}
	\acro{kB}{Kilobyte}
	\acro{MAC}{Media Access Control}
	\acro{MANET}{Mobile Ad-hoc Network}
	\acro{MB}{Megabyte}
	\acro{MDR}{Message Delivery Ratio}
	\acro{MTU}{Maximum Transmission Unit}
	\acro{NPS}{Naval Postgraduate School}
	\acro{ns-2}{Network Simulator 2}
	\acro{ns-3}{Network Simulator 3}
	\acro{OLSR}{Optimized Link State Routing}
	\acro{OMNet++}{Objective Modular Network Testbed in C++}
	\acro{ONE}{Opportunistic Network Environment}
	\acro{OPNET}{Optimum Network Performance}
	\acro{OSI}{Open Systems Interconnection}
	\acro{OSPF}{Open Shortest Path First}
	\acro{PCAP}{Packet Capture}
	\acro{PDR}{Packet Delivery Ratio}
	\acro{PRoPHET}{Probabilistic Routing Protocol}
	\acro{RFC}{Request For Comments}
	\acro{RIP}{Routing Information Protocol}
	\acro{RTT}{Round Trip Time}
	\acro{SGBR}{Social Group Based Routing Protocol}
	\acro{SUMO}{Simulation of Urban Mobility}
	\acro{SWANS}{Scalable Wireless Ad Hoc Network Simulator}
	\acro{TCP}{Transmission Control Protocol}
	\acro{TraNS}{Traffic and Network Simulation Environment}
	\acro{TTL}{Time to Live}
	\acro{USN}{\ac{US} Navy}
	\acro{UDP}{User Datagram Protocol}
	\acro{UML}{Unified Modeling Language}
	\acro{USG}{United States government}
	\acro{VANET}{Vehicular Ad-Hoc Network}
	\acro{WKT}{Well Known Text}
	\acro{XML}{Extensible Markup Language}
\end{acronym}

\begin{abstract}
We present the Epidemic routing protocol implementation in ns-3.  It is a full-featured DTN protocol in that it supports the message abstraction and store-and-haul behavior.  We compare the performance of our Epidemic routing ns-3 implementation with the existing implementation of Epidemic in the ONE simulator, and discuss the differences.
\end{abstract}

%
%

\ccsdesc[500]{Networks~Routing protocols}
\ccsdesc[500]{Computing methodologies~Model development and analysis}

\keywords{Epidemic routing, Epidemic model implementation, DTN, disruption-tolerant networking, delay-tolerant networking, MANET routing, ad-hoc networks, ns-3, network simulation}

\maketitle

\section{Introduction}
In this work we present the implementation of the Epidemic routing protocol~\cite{vahdat:2000:ERF} in ns-3~\cite{ns3:www}.  While several \ac{MANET} routing protocols are available in ns-3, it does not currently include any \ac{DTN} protocols, and development of such protocols typically proceeds with other simulators, such as The ONE simulator~\cite{Keranen:2009:OSD:1537614.1537683}.  In various routing simulation efforts~\cite{pospischil:2017:MRO,rohrer:2017:EOG,rohrer:2016:GAR} we have encountered the need to simulate both \ac{MANET}s and \ac{DTN}s in the same simulator, and so have developed an architecture for integrating the two.  Since there are no other DTN routing protocols in ns-3, we evaluate our results by comparing against the same simulation run in the \ac{ONE} simulator.


\ac{DTN} routing protocols typically deal with ``messages'' as opposed to ``packets''.  Messages may range in size from a few kB to hundreds of MB, which means they need a mechanism for fragmenting and reassembling these messages on a hop-by-hop basis to meet lower layer requirements.  \ac{DTN} routers also buffer messages if a next-hop is not available, and may replicate messages to multiple next hops.  These features differentiate \ac{DTN} routing protocols from \ac{MANET} or traditional \ac{IP} routing protocols.  This also leads to various design architectures.  \ac{DTN}s are often implemented as an application-overlay network with Convergence Layer Adapters (CLAs) for particular network types, instead of a native layer-3 routing protocol.  The application-layer approach (e.g. the Bundling Protocol~\cite{rfc5050}) has advantages in ease of development and flexibility, but prevents the overlay from interoperating with traditional IP applications.  Due to these limitations, we implement \ac{DTN} routing protocols as native ns-3 routing protocols so they can be used with the existing ns-3 application classes.  In addition we have designed an \ac{IP} convergence architecture that supports \ac{DTN} messages made up of multiple packets.  Our implementation includes application classes to generate such messages.  All of this code is available from our website\footnote{\url{https://tancad.net/projects/dtn-simulation/}}, and we are in the process of contributing it to the ns-3 distribution.

\section{Prior Work}\label{sec:Background}
This is not the first attempt at implementing the Epidemic routing protocol in ns-3.  In 2015 Alenazi \etal published their ns-3 implementation of Epidemic~\cite{Alenazi:2015:ERP:2756509.2756523}, the code for which began the review process for inclusion in ns-3 in 2013, however that process appeared to stall in 2015, and the authors did not respond to our inquiries about the status.  The existing code however was available online and we attempted to use this version for our DTN simulations, however limitations in this implementation rapidly became apparent and necessitated developing our own implementation.  Since Alenazi's implementation routes only atomic packets, not complete messages, we will refer to his implementation as PacketEpidemic in the remainder of the paper, for ease of reference.

The PacketEpidemic implementation for \ac{ns-3} implements the Epidemic logic discussed in~\cite{vahdat:2000:ERF}, however it possesses many limitations.  As mentioned, PacketEpidemic handles individual packets instead of messages.  PacketEpidemic's node discovery mechanism results in incorrect operation for nodes with large buffers, where the nodes repeatedly resend messages that have already been transferred.  Also, PacketEpidemic does not support control message fragmentation, so control messages are limited to the size of one \ac{UDP} packet.  Large message buffers can generate summary vectors that exceed the size of a \ac{UDP} packet and when this happens the \ac{ns-3} simulation crashes.  The lack of control packet fragmentation restricts the number of messages that a node can handle.  Our \ac{ns-3} Epidemic implementation addresses the need for control packet fragmentation and as a result, the control packet headers, node discovery, and data handling, are significantly different from PacketEpidemic.

\section{Code Structure}\label{ssec:CodeStruct}
\begin{figure*} [t]
\begin{tikzpicture}
	\begin{class}[font=\scriptsize,text  width=3cm]{IPv4 Routing Protocol}{0,0}
		\attribute{ }
		\operation{ }
	\end{class}
	
	\begin{class}[font=\scriptsize, text  width=3.5cm]{Routing Class}{0,-1.75}
		\implement{IPv4 Routing Protocol}
		\attribute {-  \texttt{m\_BufferSize}}
		\attribute {-  \texttt{m\_queueEntryExpireTime}}
		\attribute {-  \texttt{m\_beaconInterval}}
		\operation {+ \texttt{Recv<Protocol>}}
		\operation {+ \texttt{RouteInput}}
		\operation {+ \texttt{RouteOutput}}
		\operation {+ \texttt{SendDisjointMessages}}
		\operation {+ \texttt{SendBeacons}}
		\operation {+ \texttt{SendMessageFromQueue}}
		\operation {+ \texttt{SendACK}}
	\end{class}
	
	\begin{class}[font=\scriptsize, text  width=3.25cm]{Packet Queue Class}{8,0}
		\attribute {-  \texttt{m\_BufferSize}}
		\attribute {- \texttt{m\_queue}}
		\operation {+ \texttt{Enqueue}}
		\operation {+ \texttt{Dequeue}}
		\operation {- \texttt{Purge}}
		\operation {+ \texttt{Find}}
		\operation {- \texttt{Drop}}
		\operation {- \texttt{DropExpiredMessages}}
		\operation {+ \texttt{FindDisjointMessages}}
		\operation {+ \texttt{GetSize}}
	\end{class}
	
	\begin{class}[font=\scriptsize, text  width=3.5cm]{Queue Entry}{12,0}
		\attribute {-  \texttt{m\_packets}}
		\attribute {-  \texttt{m\_header}}
		\attribute {-  \texttt{m\_expire}}
		\attribute {-  \texttt{m\_messageID}}
		\operation {+ \texttt{AddPacket}}
		\operation {+ \texttt{GetPackets}}
		\operation {+ \texttt{SetIpv4Header}}
		\operation {+ \texttt{GetIpv4Header}}
		\operation {+ \texttt{GetExpireTime}}
		\operation {+ \texttt{SetExpireTime}}
		\operation {+ \texttt{SetMessageID}}
		\operation {+ \texttt{GetMessageID}}
		\operation {+ \texttt{GetMessageByteSize}}
		\operation {+ \texttt{GetMessagePacketTotal}}
		\operation {+ \texttt{GetCurrentPktCnt}}
		\operation {+ \texttt{GetPacketSize}}
	\end{class}
	
	\begin{class}[font=\scriptsize,text  width=3cm]{Packet Classes}{4,0}
		\attribute{}
		\operation{+ \texttt{Serialize}}
		\operation{+ \texttt{Deserialize}}
		\operation{+ \texttt{Print}}
	\end{class}
	
	\unidirectionalAssociation{Packet Classes}{}{}{Routing Class}
	\unidirectionalAssociation{Packet Classes}{}{}{Packet Queue Class}
	\unidirectionalAssociation{Packet Queue Class}{}{}{Routing Class}
	\unidirectionalAssociation{Queue Entry}{}{}{Packet Queue Class}
\end{tikzpicture}
\vspace{-12pt}
\caption{\ac{DTN} \ac{UML} Diagram}\label{fig:DTNUML}
\end{figure*}
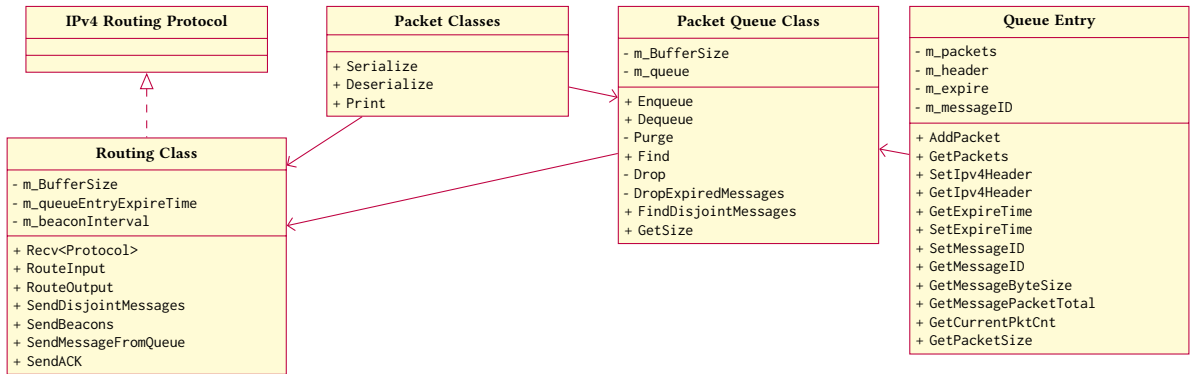

We note that implementing Epidemic routing was not our end goal, but a stepping stone to the implementation of many other DTN routing protocols in ns-3, so some design decisions take into account not just epidemic, but other protocols as well.  
We borrowed the top-level code structure from Alenazi's Epidemic~\cite{Alenazi:2015:ERP:2756509.2756523}, so all \ac{ns-3} \ac{DTN} routing protocols are broken into four main classes.  Our Epidemic implementation forwards groups of data packets called messages instead of atomic packets.  Section~\ref{sec:MessageGeneration} describes how the protocols generate messages and segment them into individual packets.  This section provides an overview of the code structure.  The \ac{DTN} routing protocol contains a group of packet classes, packet queue class, queue entry class, and routing protocol class.  The \ac{UML} diagram in Figure~\ref{fig:DTNUML} shows the relationships functions and attributes used by all \ac{ns-3} Epidemic protocols to implement \ac{DTN} logic.  We also borrow from our prior experience implementing DSDV for \ac{ns-3}~\cite{narra:2011:DSD}.

\subsection{Packet Classes}\label{sec:packetclass}
The packet classes are the packet header declarations used by the \ac{DTN} protocols.  Each packet header is its own class because the packet header is a data structure that defines a packet.  The packet classes inherit \ac{ns-3}'s \texttt{Header} class.  The \texttt{Header} class defines a packet and provides the interface for other classes to interact with the packets.  The routing class and packet queue class interact with the packet class to read and write packet headers.  Since packet headers have various types of information, they have their own accessors and mutators.  However, all packet header declarations require a \texttt{Serialize} and \texttt{Deserialize} function because of the \texttt{Header} class.  The \texttt{Serialize} function writes the header information to a packet buffer, and the \texttt{Deserialize} function reads from a packet buffer.  These functions are required because \ac{ns-3} passes packets between nodes as byte arrays.  When a node receives a packet, it uses the \texttt{Deserialize} function to obtain the header information.  The \texttt{Print} function permits another class to print the header to the screen or log file.

\subsection{Packet Queue and Queue Entry Classes}\label{sec:packetqueue}
The Packet Queue Class manages a node's message buffer.  The Routing Protocol class interacts with the Packet Queue class to manage messages and generate control packets.  The Packet Queue class implements a protocol's buffer management scheme, something which is much more significant in other DTN protocols.  The \texttt{m\_BufferSize} attribute defines the maximum size of the message buffer in bytes.  \texttt{Enqueue} adds messages to the buffer, and \texttt{Dequeue} removes messages from the buffer.  After a node adds a message to the message buffer, \texttt{DropExpiredMessages} removes expired messages.  If the message buffer is full, \texttt{Purge} removes messages according to the protocol's queue management scheme.  \texttt{FindDisjointMessages} generates the list of messages to replicate according to a protocol's message priority.  The \texttt{Drop} function removes a selected message from the buffer.  The \texttt{GetSize} function returns the number of messages in the buffer.

Since messages are groups of packets, the message buffer requires a data structure to group packets.  The \texttt{m\_queue} is a map matching a message ID to a queue entry defined by the Queue Entry class.  A queue entry is a data structure that stores the packets belonging to a message, the \ac{IP} header, expiration time, and message ID.  The \texttt{GetMessageByteSize} returns the number of packets contained in message.  The \texttt{GetMessagePacketTotal} returns the total number of packets belonging to a message.  The \texttt{GetCurrentPktCnt} returns the number of packets currently contained in the queue entry.  The \texttt{GetPacketSize} returns the size a data packet in a message.  \texttt{AddPacket} adds a packet to the queue entry.  \texttt{GetPackets} returns all of the packets contained in a message.  The Packet Queue uses the Queue Entry functions to generate control packets, manage the message buffer, and retrieve messages for the Routing Protocol class.

\subsection{Routing Protocol Class}\label{sec:routingclass}
The Routing Protocol class inherits from \ac{ns-3}'s \texttt{Ipv4RoutingProtocol}.  The Routing Protocol class implements the control logic of the \ac{DTN} protocols.  The \texttt{Recv<Protocol>} function executes the control packet exchange based on the packet headers defined in the packet class.  The Routing Class initializes the packet queue.  The routing class interacts with the packet queue class to generate control packets, and the routing protocol class interacts with the packet queue class to store and retrieve messages.  The \texttt{SendBeacons} function transmits beacon packets at the specific interval.  The \texttt{SendDisjointMessages} calls the \texttt{FindDisjointMessages} from the Packet Queue class to generate the list of message to transmit to another node.  Then \texttt{SendDisjointMessages} calls \texttt{SendMessageFromQueue} to transmit the messages from the generated message list.  When a node receives a packet, \texttt{RouteInput} determines the interface, buffer, or function to execute.  \texttt{RouteInput} handles the logic for buffering incomplete messages and acknowledging messages described in Section~\ref{sec:MessageGeneration}.  \texttt{RouteOutput} handles packets leaving a node.

\section{Message Generation and Handling}\label{sec:MessageGeneration}
The \ac{ONE} and \ac{ns-3} handle node data differently.  This section discusses how \ac{ns-3} and the \ac{ONE} implement message handling and generation.  The discussion includes the differences between the \ac{ns-3} PacketEpidemic implementation and our \ac{ns-3} Epidemic implementation to handle messages, and \ac{ns-3} application-layer message-traffic generation.

\subsection{\ac{ONE} Behavior}
In the \ac{ONE}, traffic generation uses messages.  The \ac{ONE} does not have packets like \ac{IP} networks, so the \ac{ONE} does not have packet header definitions.  Messages are similar to bundles from the Bundling Protocol~\cite{rfc5050} because messages are the base unit of \ac{DTN} data.  Unlike \ac{IP} packets, messages can be any size.  The \ac{ONE} handles messages as a single object.  The \ac{ONE} does not fragment messages, and the \ac{ONE} does not permit partial messages to propagate throughout the network.  If a node does not completely receive a message, then the node drops the partially received message.  In the \ac{ONE}, messages do not carry control instructions, and nodes share control information by directly accessing another node's data structures in memory.  The \ac{ONE} does not include the overhead of exchanging control instructions in its simulation results.

\subsection{\ac{ns-3} Behavior}
Since \ac{ns-3} implements the entire network stack, our \ac{ns-3} protocols define groups of packets generated by one source node destined to another node as a message.  A message is equivalent to RFC-5050's bundles as the base unit of \ac{DTN} data.  The \ac{UDP} packets used to share routing information between nodes are control packets. Control packets are not messages because they are routing protocol specific, so they are not the base unit of \ac{DTN} data. Our protocols assume an \ac{IP}-based convergence layer.  Unlike the \ac{ONE}, \ac{ns-3} cannot generate a message as a single object of any size, so large messages are segmented into groups of individual packets.  Each packet is assigned a header with a custom identifier that associates each packet with the rest of the \ac{DTN} message.  This architecture integrates with the existing code base, and does not require the modification of protocols below the routing layer.  
The following subsections discuss how \ac{ns-3} defines and handles messages.  The discussion includes the differences from PacketEpidemic message handling, and our \ac{ns-3} message-traffic generation.
 
\subsection{Message Definition}
Our \ac{ns-3} \ac{DTN} data packet header shown in Figure~\ref{fig:MsgHdr} is the custom header that identifies a packet belonging to a message.  The Bundling Protocol (\ac{RFC} 5050)~\cite{rfc5050} influenced our \ac{ns-3} \ac{DTN} data packet header design, but our \ac{ns-3} \ac{DTN} protocols do not implement the Bundling Protocol.  The PacketEpidemic implementation does not have a header for \ac{DTN} message layer, instead, it uses only \ac{UDP} packets as the base unit of data with the \ac{IP} address identifying source and destination nodes.

\begin{figure}[t]
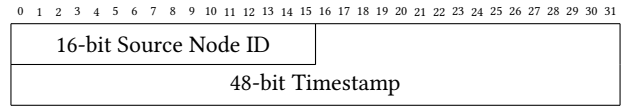

	\centering
	\begin{bytefield}[bitwidth=.8em]{32}
		\bitheader{0-31} \\
		\bitbox{16}{16-bit Source Node ID} & \bitbox[lrt]{16}{} \\
		\wordbox[lrb]{1}{48-bit Timestamp}\\
	\end{bytefield}
	\vspace{-12pt}
	\caption{Message Identification Number}\label{fig:MsgID}
\end{figure}

Each \ac{DTN} data packet header contains a Message Identification Number, last hop, packet count, and packet index.  The Message Identification Number in Figure~\ref{fig:MsgID} is a 64-bit unsigned integer derived from the source Node Identification Number and timestamp.  The first 16-bits of the Message Identification Number are the source Node Identification Number.  The last 48-bits are the message's generation timestamp in microseconds.  The source Node Identification Number used by the \ac{DTN} packet header is the node number from \ac{ns-3}, which automatically assigns a unique integer to every node in the simulation.

\newcommand{\baselinealign}[1]{%
  \centering
  \strut\small#1%
}
\newcommand{\colorbitbox}[3]{%
  \sbox0{\bitbox{#2}{#3}}%
  \makebox[0pt][l]{\textcolor{#1}{\rule[-\dp0]{\wd0}{\ht0}}}%
  \bitbox{#2}{#3}%
}

\begin{figure}[t]
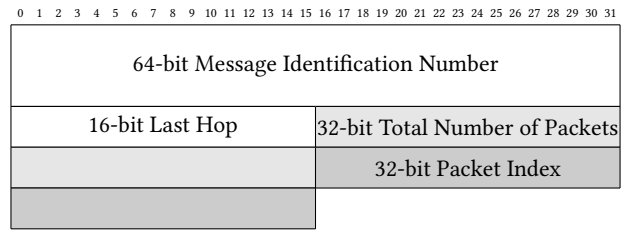

	\centering
	\begin{bytefield}[bitwidth=.8em]{32}
		\bitheader{0-31} \\
		\wordbox{2}{64-bit Message Identification Number}\\
		\bitbox{16}{16-bit Last Hop} & \colorbitbox{gray!20}{16}{32-bit Total Number of Packets}\\
		\colorbitbox{gray!20}{16}{} & \colorbitbox{gray!40}{16}{32-bit Packet Index}\\
		\colorbitbox{gray!40}{16}{}\\
	\end{bytefield}
	\vspace{-12pt}
	\caption{\ac{DTN} Data Packet Header}\label{fig:MsgHdr}
\end{figure}

The range of scenarios we expect to simulate require a node to generate several messages in a millisecond at most, but will not require generation of more than one message in a microsecond.  The timestamp is in microseconds to ensure that every Message Identification Number is unique.  While a per-node sequential message counter would also create a unique identifier, our \ac{ns-3} protocols require message generation time and source node identification for routing decisions, so this information is dual-purpose.  Some scenarios studied would exceed the largest time in microseconds represented by a 32-bit timestamp, therefore we use 48-bits for the timestamp.  The last hop field is the 16-bit Node Identification Number of the last node that forwarded the message.  Nodes use the last hop to buffer incomplete messages for message reconstruction.  The 32-bit total number of packets and the 32-bit packet index support reordering during message reassembly.

\subsection{Message Handling}
In PacketEpidemic, nodes pushed all packets selected for transmission to the link layer immediately after completing the control packet exchange.  \ac{ns-3}'s link layer contains a packet queue that has a limited size.  This link-layer packet queue is \ac{FIFO} and cannot be manipulated by higher layers in the network stack, they can only add packets~\cite{NS3Manual}.  When a node's message buffer equals or exceeds the size of the packet queue, then the node will fill the packet queue and any new packets sent to the full queue are dropped.  The link layer will attempt to transmit the packets in \ac{FIFO} order regardless if the destination node is connected.  If the first node moves out of range, then the link layer will still transmit the packets.  As a result, the node wastes available bandwidth and meeting opportunities.  This behavior is not a requirement of the Epidemic routing protocol, but is a limitation specific to the PacketEpidemic implementation.

\begin{figure}[t]
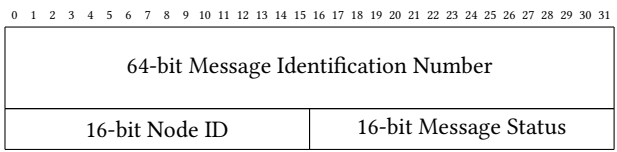

	\centering
	\begin{bytefield}[bitwidth=.8em]{32}
		\bitheader{0-31} \\
		\wordbox{2}{64-bit Message Identification Number} \\
		\bitbox{16}{16-bit Node ID} & \bitbox{16}{16-bit Message Status} \\
	\end{bytefield}
	\vspace{-12pt}
	\caption{\ac{DTN} Acknowledgement Header}\label{fig:DTNAck}
\end{figure}

In order to improve link utilization, \texttt{ACK} packets in Figure~\ref{fig:DTNAck} control the message exchange sequence.  \texttt{ACK} packets do not acknowledge messages reaching their final destination, rather, they are a hop-by-hop acknowledgement of messages transmitted between two connected nodes.  When a node receives a complete message from another node, it sends an acknowledgement packet consisting of the 64-bit Message Identification Number, 16-bit Node Identification Number, and 16-bit Message Status.  The Message Identification Number is the received message's message ID.  The Node Identification Number is the node that received the message.  The Message Status block permits adding reliability in future work.  The Message Status indicates if a node received a message successfully.  If the message transfer is unsuccessful, then Message Status is zero.  We do not retransmit such messages, but such reliability may be a desirable future enhancement and is supported by the header.

After the transmitting node determines message priority, the node fetches the first message.  A node transmits the complete message and waits for an \texttt{ACK} before sending the next message.  The receiving node queues the message's packets in a message reception buffer.  The receiving node maintains a message reception buffer for each node that is connected.  Each connection's receiving buffer can buffer only one message.  When a message is complete, the receiving node transmits an \texttt{ACK}. \texttt{RouteInput} handles message reconstruction and \texttt{ACK} generation.  Upon receiving the \texttt{ACK}, the transmitting node transmits the next message.  The \texttt{Recv<Protocol>} function handles \texttt{ACK} reception.  If the connection breaks, then the node resets the message reception buffer for that connection.  A node considers a connection broken when the node does not receive any packets from that neighbor for two beacon intervals.  When a node does not receive a complete message, it deletes the partial message.  As a result, nodes do not forward partial messages.  

Deleting partial messages may seem wasteful, but dropped messages due to a lost packet occurs infrequently (\eg less than 0.1\% of the time in the Helsinki scenario).

While \texttt{ACK}s improve link efficiency, they also have limitations.  First, a node requires a large link layer buffer because the buffer must store all of the packets contained in a message.  When the link layer's packet buffer exceeds the packet limit, the node removes the newest packets from the queue.  As a result, those packets fail to transfer causing messages to drop.  \ac{ns-3}'s link layer limits the number of packets that can be stored in the buffer, so the scenario configuration file must increase the link layer's buffer to accommodate the expected message sizes.  Second, the link layer buffer deletes packets based on a time-limit.  When a packet exceeds the time-limit, the link layer buffer deletes the packet.  For the purposes of this work, the example scenarios set the link layer buffer to the size of the node's message buffer.  The scenarios set the packet queue time-limit to two beacon intervals because two beacon intervals correspond to a dropped connection.

\subsection{Message Generation}
The \ac{DTN} Application generates message traffic using the \ac{DTN} Packet Header, and is adapted from the On-Off Application.  The \ac{DTN} Application generates \ac{UDP} packets with the \ac{DTN} Packet Header.  
The message size parameter in bytes and packet size determines the number of generated packets.  The \ac{DTN} Application generates one message according to the entered parameters.  The Message Identification Number uses the source Node Identification Number and message generation time.  If a scenario requires more than one message, then the scenario must include multiple \ac{DTN} Application message generators.

Since future work may use packets instead of messages for data traffic, our \ac{ns-3} \ac{DTN} protocol implementation is backwards compatible with raw \ac{UDP} packets.  When a node generates a \ac{UDP} packet that does not have the \ac{DTN} Packet Header, the DTN router on that node creates a \ac{DTN} Packet Header for that data packet.  \texttt{RouteInput} handles \ac{DTN} Packet Header generation for standard \ac{UDP} packets.  The Message Identification Number uses the packet's generation time in microseconds and the packet's source Node Identification Number to generate the Message Identification Number.  A node then treats each packet as an individual message, permiting \ac{ns-3}'s default \ac{UDP} packet generators to work with the \ac{DTN} protocols.

\section{Node Discovery}\label{sec:NodeDiscovery}
DTNs do not have constant connectivity, so nodes must discover other nodes to initiate a connection.  Unlike \ac{ns-3}, the \ac{ONE} simulator handles node discovery independently of any routing protocol.  As a result, the \ac{ns-3} DTN protocols must include a node discovery mechanism.  

In PacketEpidemic, nodes transmit \texttt{BEACON} control packets at a specific time interval for node discovery.  The \texttt{BeaconInterval} defines the frequency of beacon transmission.  To ensure that all nearby nodes can receive the beacon, the node broadcasts a beacon using the network's broadcast address.  Nodes are likely to be synchronized in sending broadcasts, resulting in lost beacons due to collisions.  In order to prevent synchronization among nodes, a uniform random variable staggers the beacons.  The \texttt{BeaconRandomness} variable defines the upper bound of the uniform random distribution to add to the base beacon interval.  Since each beacon starts the exchange process between nodes, nodes with many packets buffered may already be exchanging packets when the next beacon interval occurs.  The \texttt{HostRecentPeriod} prevents hosts from re-exchanging redundant control packets~\cite{Alenazi:2015:ERP:2756509.2756523}, however nodes with large buffers will exceed the \texttt{HostRecentPeriod} used PacketEpidemic resulting in the nodes re-exchanging control packets and messages while already transferring messages.  This causes messages to be sent and re-sent multiple times wasting bandwidth.

Our Epidemic implementation also uses beacons for node discovery.  A node broadcasts beacons at a set interval plus a random delay to minimize beacon collisions.  However, we do not use a \texttt{HostRecentPeriod} to prevent hosts from re-exchanging redundant control packets, instead each node remembers the neighbors with which it is currently in contact.  Every time a node receives a control packet or data packet from another node; it updates its record of connected nodes.  A node considers a connection broken when the node does not receive anything from the neighbor for two beacon intervals.

A node can have more than one radio interface and \ac{IP} address, so \ac{IP} address is not a sufficient identifier to prevent two nodes from restarting a connection when they detect the second radio interface.  

The \texttt{BEACON} control packet is a \texttt{MessageType} header with the field set to \texttt{BEACON}.  The \texttt{MessageType} header field indicates the type of control packet.  Since a node does not know whether another node has more than one \ac{IP} address, the beacon must include the Node Identification Number.  Therefore, the \texttt{MessageType} header includes the 16-bit Node Identification Number.  Figure~\ref{fig:DTNMtype} illustrates the \texttt{MessageType} Header.  \ac{ns-3} provides a unique number to every node in a simulation.  A node uses the Node Identification Number to determine whether a node is considered connected.  If a node is not connected, then the node adds the Node Identification Number with a timestamp to its list of connected nodes.  Then, the node continues the control packet exchange.  If the Node Identification Number is in the list of connected nodes and the connected timestamp does not exceed two beacon intervals, then the node ignores the \texttt{BEACON}.  However, every data packet and control packet other than \texttt{BEACON}s update the connection timestamp.  A \texttt{BEACON} changes the timestamp only when the timestamp exceeds two beacon intervals because it is considered as a new connection.

\begin{figure}[t]
	\centering
	\begin{bytefield}[bitwidth=.8em]{24}
        		\bitheader{0-23} \\
		\bitbox{8}{Message Type} & \bitbox{16}{16-bit Node ID} \\
	\end{bytefield}
	\vspace{-12pt}
	\caption{\ac{DTN} \texttt{MessageType} Header}\label{fig:DTNMtype}
\end{figure}

\section{Epidemic Logic}\label{sec:EpidemicDTN}
\begin{figure}[t]
	\centering
	\begin{tikzpicture}
		\draw [black, thick] (0,0) circle [radius=1.25];
		\node [scale=5] at (0,0) {A};
		\draw [black, thick] (6,0) circle [radius=1.25];
		\node [scale=5] at (6,0) {B};
		\draw [<->, thick] (.9,1.0) -- (5.1,1.0);
		\node at (3,1.25) {\texttt{BEACON}};
		\draw [->, thick] (1.25,0.33) -- (4.75,0.33);
		\node at (3,0.58) {\texttt{REPLY}};
		\draw [<-, thick] (1.25,-0.33) -- (4.75,-0.33);
		\node at (3,-0.08) {\texttt{REPLY\_BACK}};
		\draw [<->, thick, dashed] (.9,-1.0) -- (5.1,-1.0);
		\node at (3,-0.75) {Messages};
	\end{tikzpicture}
	\vspace{-12pt}
	\caption{Epidemic Control Packet Exchange Sequence}\label{fig:EpidemicDTNConM}
\end{figure}
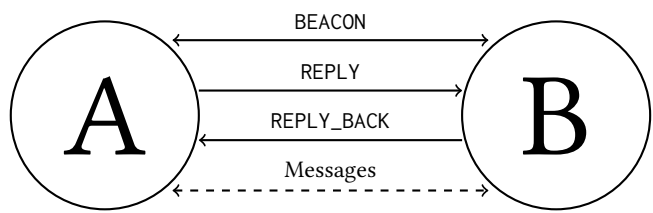

The \ac{ns-3} Epidemic control logic performs four main steps as shown in Figure~\ref{fig:EpidemicDTNConM}.  After receiving a \texttt{BEACON}, nodes exchange \texttt{REPLY} and \texttt{REPLY\_BACK} control packets.  \texttt{REPLY} and \texttt{REPLY\_BACK} control packets are the summary vectors discussed in~\cite{vahdat:2000:ERF}.  The goal of sending message summaries is to avoid sending messages that the other node already contains in its buffer.  Since both nodes are likely to send a response to a beacon at the same time, an anti-entropy session prevents node responses from colliding.  The node with the lower \ac{IP} address sends its message summary first as a \texttt{REPLY} packet.  When the node with the higher \ac{IP} address receives the \texttt{REPLY} packet, it sends a response using the \texttt{REPLY\_BACK} packet.  After a node receives the list of messages that the other node contains, the node sends messages that the other node does not have in its buffer.

\subsection{Message Identification and Limits}
Since our implementation manages messages instead of packets, we use a different \texttt{EpidemicHeader} than PacketEpidemic~\cite{Alenazi:2015:ERP:2756509.2756523}.  Figure~\ref{fig:EpidemicDTNEpcHdr} illustrates the \texttt{EpidemicHeader}.  First, the 64-bit Message Identification Number from Section~\ref{sec:MessageGeneration} replaced the 32-bit packet identification number.  Since the Message Identification Number includes the source node and timestamp of message generation, we do not use a separate timestamp field.

\begin{figure}[t]
	\centering
	\begin{bytefield}[bitwidth=.8em]{32}
        		\bitheader{0-31} \\
        		\wordbox{2}{64-bit Message Identification Number} \\
        		\bitbox{32}{32-bit Hop Count} \\
	\end{bytefield}
	\vspace{-12pt}
	\caption{\texttt{EpidemicHeader}}\label{fig:EpidemicDTNEpcHdr}
\end{figure}

The \ac{ONE} uses hop limits and message \ac{TTL} to reduce network resource consumption.  The lower 48-bit portion of the Message Identification Number permits the routing protocol to remove expired messages from a node's buffer.  When a node receives a message, the node checks the timestamp against the maximum age of a message.  If the timestamp exceeds the amount of time a message can live, then the node discards the message.  The \texttt{EpidemicHeader} 32-bit hop count field is used limit the number of hops that a message can traverse.  When nodes generate a message, they initialize the hop count to the maximum number of hops allowed.  At each hop, the node decrements the hop count and when the hop count reaches zero, the message is discarded.

\subsection{Control Packet Identification}
The \texttt{MessageType} Header in Figure~\ref{fig:DTNMtype} identifies control packets.  Since the \ac{ONE} simulator uses shared data structures to exchange routing information it did not implement control packets.  In our implementation, control packets are not messages (the do not have a DTN message header).  Control packets share routing specific information between two neighboring nodes.  The 8-bit Message Type field indicates the type of control packet.  For Epidemic, control packets are \texttt{BEACON}, \texttt{ACK}, \texttt{REPLY}, and \texttt{REPLY\_BACK}.  The \texttt{MessageType} Header encapsulates the control packets in order to identify the control packet.  When a node receives a packet, it checks the Message Type field for the packet type to call the appropriate packet header class.

\subsection{Summary Vector}
While Epidemic uses the same control packet sequence as PacketEpidemic, the \texttt{SummaryVectorHeader} in Figure~\ref{fig:EpidemicDTNSMVHdr} differs.  The \texttt{SummaryVectorHeader} defines \texttt{REPLY} and \texttt{REPLY\_BACK} control packets.  We use a 64-bit Message Identification Number in place of packetEpidemic's 32-bit packet identification number.  PacketEpidemic's 32-bit Summary Vector Length counts all of the packets held by a node's buffer.  In our Epidemic implementation, the 16-bit Summary Vector Length counts the number of Message Identification Numbers in the \texttt{SummaryVectorHeader}.  A 16-bit unsigned integer is large enough cover a \texttt{SummaryVectorHeader}.  Since the \texttt{MessageType} Header encapsulates the \texttt{SummaryVectorHeader}, a node identifies the other node using the Node Identification Number instead of the \ac{IP} address because nodes can have more than one \ac{IP} address.

\begin{figure}[t]
	\centering
	\begin{bytefield}[bitwidth=.8em]{32}
		\bitheader{0-31} \\
		\bitbox{16}{16-bit Fragmentation Block} & \bitbox{16}{16-bit Summary Vector Length} \\
		\wordbox{2}{64-bit Message Identification Number \#1} \\
		\wordbox{2}{. . .} \\
		\wordbox{2}{64-bit Message Identification Number \#N} \\
	\end{bytefield}
	\vspace{-12pt}
	\caption{Epidemic \texttt{SummaryVectorHeader}}\label{fig:EpidemicDTNSMVHdr}
\end{figure}

PacketEpidemic did not support control packet fragmentation.  In order to support fragmentation, the Fragmentation Block identifies whether there are more \texttt{SummaryVectorHeader} packets.  When the Fragmentation Block is one, more \texttt{SummaryVectorHeader} packets remain.  When the Fragmentation Block is zero, that packet is the last \texttt{SummaryVectorHeader}.  Once a node receives one with the Fragmentation Block set to zero, the node continues the message exchange sequence.  The control packet fragmentation can tolerate packet loss if the lost packet had the fragmentation block set to one, but the fragmentation protocol does not support retransmission of control packets.  If the lost control packet had the fragmentation block set to zero, then the control packet sequence would stop until a beacon restarts the exchange sequence.

\begin{table*}[t]
	\centering
	\caption{Helsinki Simulator Performance Difference}
	\label{tab:HelEpdPerfDiff}
	\resizebox{\textwidth}{!}{
	\begin{tabular}{| l | c | c | c | c | c | c | c | c | c | c | c | c | c | c | c |}\hline
		 & \multicolumn{5}{ c |}{\textbf{MDR}} & \multicolumn{5}{ c |}{\textbf{Average Latency}} & \multicolumn{5}{ c |}{\textbf{Message Replication Overhead Ratio}} \\ \hline
		 & \textbf{5 MB} & \textbf{10 MB} & \textbf{25 MB} & \textbf{50 MB} & \textbf{100 MB} & \textbf{5 MB} & \textbf{10 MB} & \textbf{25 MB} & \textbf{50 MB} & \textbf{100 MB} & \textbf{5 MB} & \textbf{10 MB} & \textbf{25 MB} & \textbf{50 MB} & \textbf{100 MB} \\ \hline
		 \textbf{6 Mbps}   & -62\% & -70\% & -78\% & -78\% & -77\% & -2.8\% & 27\% & 72\% & 82\% & 73\% & -27\% & -11\% & 50\% & 35\% & 43\% \\ \hline
		 \textbf{12 Mbps} & -32\% & -45\% & -56\% & -62\% & -61\% & -27\% & -5.3\% & 48\% & 82\% & 99\% & -56\% & -50\% & -27\% & -15\% & -14\% \\ \hline
		 \textbf{24 Mbps} & 11\% & -10\% & -25\% & -31\% & -36\% & -38\% & -23\% & 23\% & 66\% & 105\% & -69\% & -68\% & -59\% & -55\% & -49\% \\ \hline
		 \textbf{36 Mbps} & 34\% & 12\% & -2.7\% & -12\% & -20\% & -44\% & -29\% & 12\% & 56\% & 95\% & -73\% & -73\% & -69\% & -67\% & -62\% \\ \hline
		 \textbf{54 Mbps} & 50\% & 38\% & 17\% & 8.9\% & -4.2\% & -47\% & -32\% & 6.8\% & 42\% & 80\% & -76\% & -77\% & -76\% & -76\% & -72\% \\ \hline
	\end{tabular}}
\end{table*}

\begin{table*}[t]
	\centering
	\caption{Bold Alligator Simulator Performance Difference}
	\label{tab:BAEpdPerfDiff}
	\resizebox{\textwidth}{!}{
	\begin{tabular}{| l | c | c | c | c | c | c | c | c | c | c | c | c | }\hline
		 & \multicolumn{4}{ c |}{\textbf{MDR}} & \multicolumn{4}{ c |}{\textbf{Average Latency}} & \multicolumn{4}{ c |}{\textbf{Message Replication Overhead Ratio}} \\ \hline
		 & \textbf{5 MB} & \textbf{10 MB} & \textbf{25 MB} & \textbf{50 MB} & \textbf{5 MB} & \textbf{10 MB} & \textbf{25 MB} & \textbf{50 MB} & \textbf{5 MB} & \textbf{10 MB} & \textbf{25 MB} & \textbf{50 MB} \\ \hline
		 \textbf{12 Mbps} & -77\% & -74\% & -73\% & -74\% & 60\% & 66\% & 86\% & 90\% & -19\% & -25\% & -26\% & -27\% \\ \hline
		 \textbf{24 Mbps} & -74\% & -71\% & -69\% & -70\% & 58\% & 56\% & 96\% & 99\% & -21\% & -28\% & -29\% & -29\% \\ \hline
		 \textbf{36 Mbps} & -72\% & -70\% & -67\% & -68\% & 52\% & 60\% & 92\% & 108\% & -21\% & -32\% & -29\% & -31\% \\ \hline
		 \textbf{54 Mbps} & -71\% & -69\% & -66\% & -67\% & 53\% & 60\% & 89\% & 100\% & -30\% & -32\% & -31\% & -31\% \\ \hline
	\end{tabular}}
\end{table*}

\begin{table*}[t]
	\centering
	\caption{Omaha Simulator Performance Difference}
	\label{tab:OmEpdPerfDiff}
	\resizebox{\textwidth}{!}{
	\begin{tabular}{| l | c | c | c | c | c | c | c | c | c | c | c | c | c | c | c |}\hline
		 & \multicolumn{5}{ c |}{\textbf{MDR}} & \multicolumn{5}{ c |}{\textbf{Average Latency}} & \multicolumn{5}{ c |}{\textbf{Message Replication Overhead Ratio}} \\ \hline
		 & \textbf{5 MB} & \textbf{10 MB} & \textbf{25 MB} & \textbf{50 MB} & \textbf{100 MB} & \textbf{5 MB} & \textbf{10 MB} & \textbf{25 MB} & \textbf{50 MB} & \textbf{100 MB} & \textbf{5 MB} & \textbf{10 MB} & \textbf{25 MB} & \textbf{50 MB} & \textbf{100 MB} \\ \hline
		 \textbf{6 Mbps}   & 0\% & -16\% & -25\% & -31\% & -35\% & -27\% & -17\% & 8.8\% & 26\% & 43\% & -74\% & -73\% & -60\% & -61\% & -65\% \\ \hline
		 \textbf{12 Mbps} & 10\% & -8.1\% & -15\% & -21\% & -26\% & -35\% & -22\% & 0\% & 13\% & 33\% & -77\% & -75\% & -65\% & -68\% & -71\% \\ \hline
		 \textbf{24 Mbps} & 19\% & 0\% & -15\% & -16\% & -20\% & -39\% & -30\% & -10\% & 3.7\% & 29\% & -79\% & -79\% & -69\% & -74\% & -75\% \\ \hline
		 \textbf{36 Mbps} & 23\% & 0\% & -9.1\% & -11\% & -16\% & -42\% & -29\% & 12\% & 5.7\% & 29\% & -80\% & -79\% & -72\% & -75\% & -78\% \\ \hline
		 \textbf{54 Mbps} & 29\% & 1.9\% & -8.8\% & -13\% & -17\% & -41\% & -30\% & -9.1\% & 3.2\% & 26\% & -83\% & -81\% & -75\% & -78\% & -79\% \\ \hline
	\end{tabular}}
\end{table*}

\section{Evaluation}
Our primary interest in implementing DTN protocols in ns-3 is to gain fidelity and account for the cost associated with control message exchanges and the overhead of layer-2 protocols, which can not be simulated in the \ac{ONE} and other popular DTN simulators.  With this in mind, our reference point is the performance of the same protocol in the \ac{ONE} simulator.  We note that this comparison would be impossible without support for segmenting messages into multiple packets, due to the use of relatively large messages as the base unit of transmission in the \ac{ONE} and other DTN simulators.  We compare the protocols using three different mobility scenarios, all of which are run in the \ac{ONE}, which outputs and ns-2 mobility trace which in turn is input to ns-3.  In this way, while there is still randomness in the mobility, there is no difference in the mobility of the nodes between the two simulators.  Each parameter set is run 10 times with different seeds for the random number generator.  The Helsinki mobility scenario is an urban environment, which is the default scenario for the \ac{ONE} simulator and hence appears commonly in DTN simulation literature.  Bold Alligator is our interpretation of a US Marine Corps exercise, and Omaha is our interpretation of the historic amphibious assault at Omaha Beach during World War II.  While we do not have room here to fully specify each scenario, they are presented in detail in LT Mauldin's master's thesis~\cite{mauldin:2017:CAO}.

\subsection{Helsinki}
Table~\ref{tab:HelEpdPerfDiff} shows that \ac{ns-3}'s Epidemic returns lower \ac{MDR}, higher average latency, and lower message replication overhead.  Both simulators show that Epidemic's \ac{MDR} increases as buffer size increases.  Link layer overhead and control packets reduce available bandwidth to share messages.  The \ac{ONE} does not include the time or bandwidth consumed by control packets, so the \ac{ONE} version increases message replication because nodes have more time to share messages.  \ac{ns-3}'s increased sensitivity to transmission speed and higher latency highlights the impact of the added overhead.

\subsection{Bold Alligator}
In \ac{ns-3} and the \ac{ONE}, Epidemic's \ac{MDR} and average latency increase asymptotically with buffer size.  Epidemic's average hop count increases from the 5 \ac{MB} to 10 \ac{MB} buffer, but Epidemic shows minimal change in average hop count for larger message buffers in both simulators.  Message replication overhead in the \ac{ONE} and \ac{ns-3} follow the same trend.  Larger message buffers return minimal change in message replication overhead.

While both versions of Epidemic share trends, \ac{ns-3} consistently returns lower \ac{MDR}, higher average latency, and lower message replication overhead.  To illustrate this observation, Table~\ref{tab:BAEpdPerfDiff} contains the percent difference between \ac{ns-3} and the \ac{ONE}.  \ac{ns-3}'s \ac{MDR} is 66\% to 74\% percent lower than the \ac{ONE}.  Average latency is 52\% to 108\% higher.  As buffer size increases, the difference in average latency between \ac{ns-3} and the \ac{ONE} increases.  Message replication overhead is 19\% to 31\% lower in \ac{ns-3}.  Larger message buffers and higher speed radios increase the message replication overhead performance gap between simulators.

\subsection{Omaha}
Epidemic's \ac{MDR} and average latency increase asymptotically with buffer size.  Message replication overhead decreases with larger message buffers, the \ac{ONE} shows a steep decrease compared to \ac{ns-3}'s shallow decrease in message replication overhead.  Both versions of Epidemic show that average latency increases asymptotically with larger message buffers.  Epidemic's performance in \ac{ns-3} is close to the \ac{ONE} in Table~\ref{tab:OmEpdPerfDiff}.  \ac{ns-3}'s \ac{MDR} is within 35\% of the \ac{ONE}, and the simulators match at several data points.  At small buffer sizes, \ac{ns-3} has lower latency than the \ac{ONE}, but large message buffers show \ac{ns-3} has higher latency.  With respect of message replication overhead, \ac{ns-3}'s message replication overhead is 60\% to 83\% lower than the \ac{ONE}.  Higher transmission speeds increase the performance gap in message replication overhead between \ac{ns-3} and the \ac{ONE}. 

\begin{figure*}[t]
  \centering
  \subfigure[Aggregate Message Delivery Ratio]{
	\includegraphics[width=.32\linewidth]{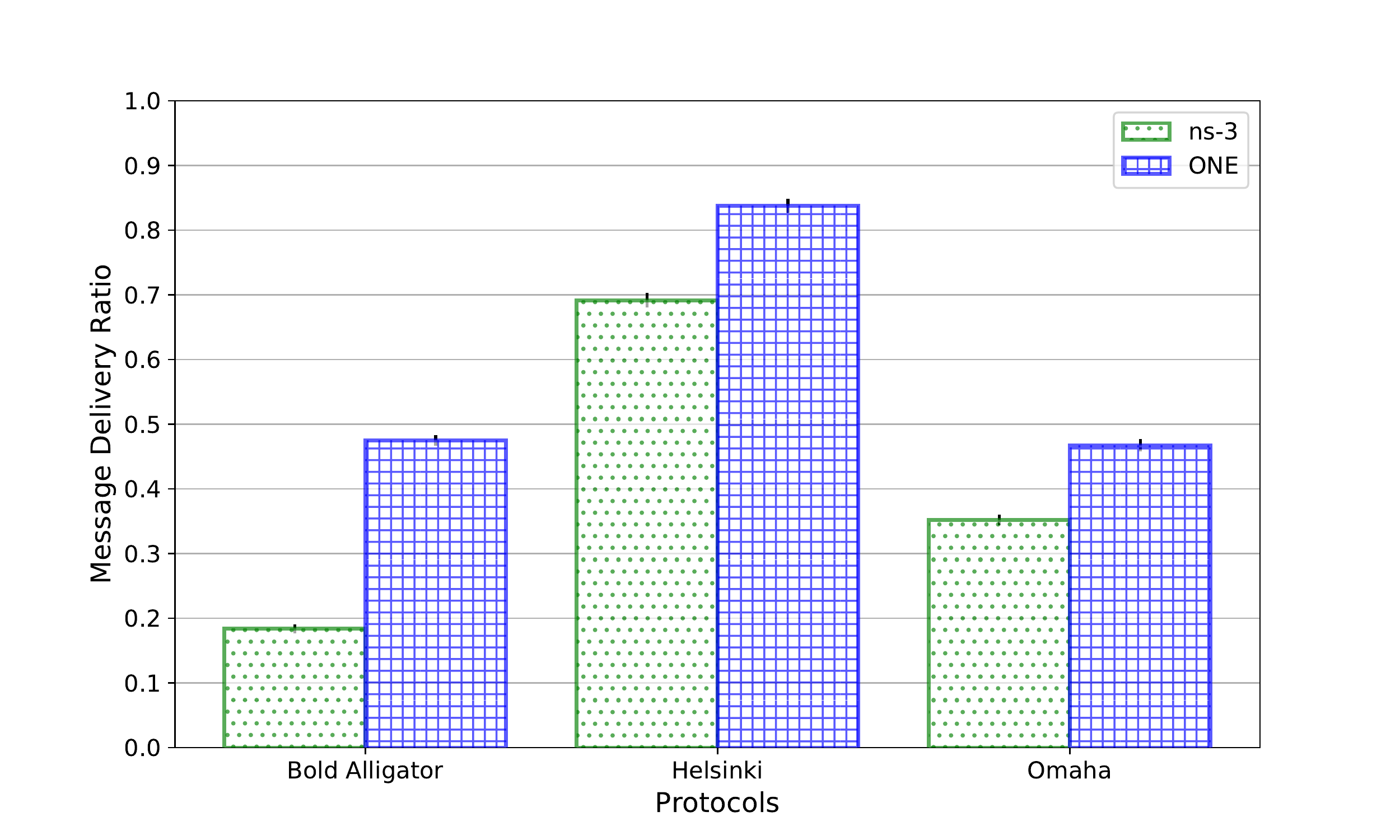}
	\label{fig:SimAggMDR}
	}
\hfill
  \subfigure[Aggregate Average Latency]{
	\includegraphics[width=.32\linewidth]{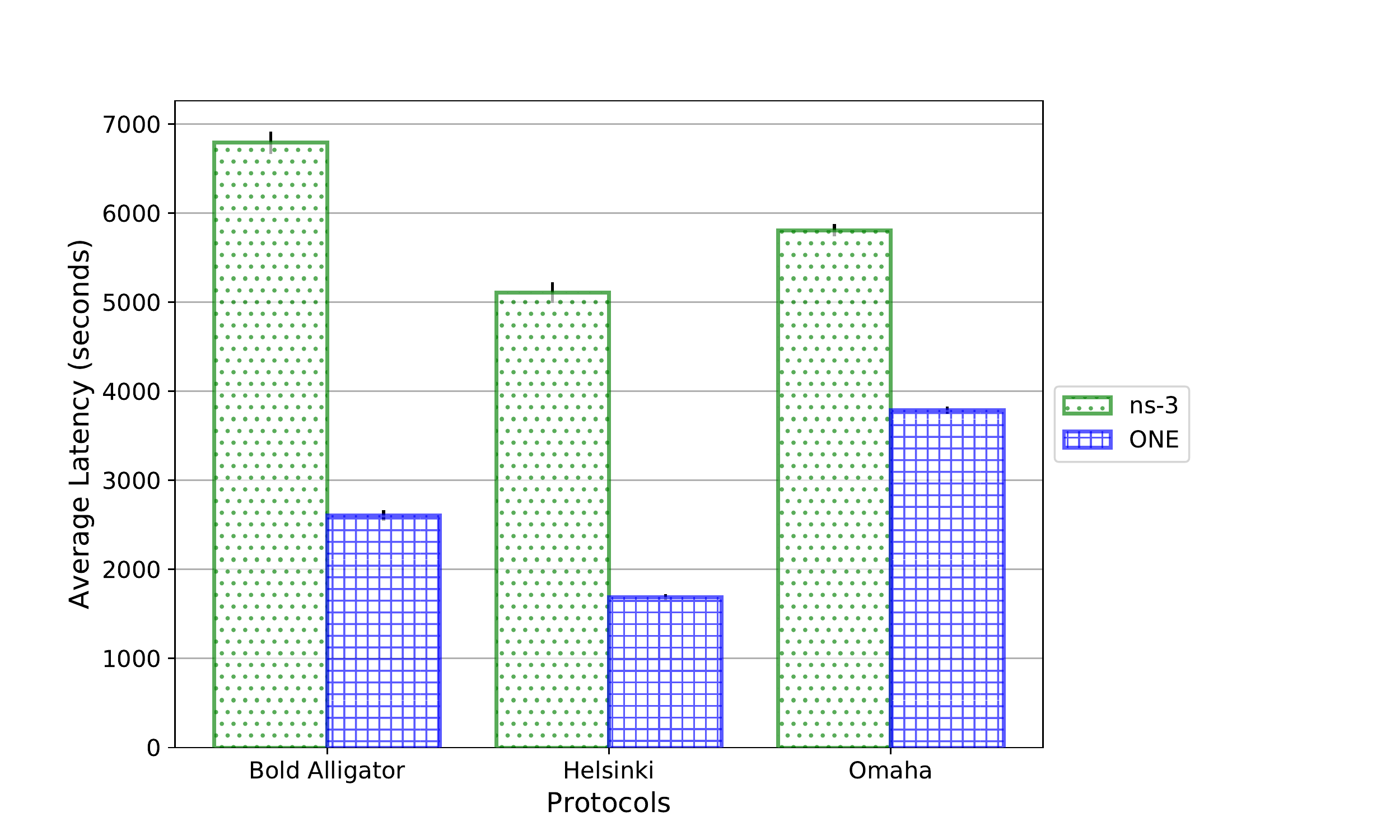}
	\label{fig:SimAggLat}
	}
\hfill
  \subfigure[Aggregate Message Replication Overhead]{
	\includegraphics[width=.32\linewidth]{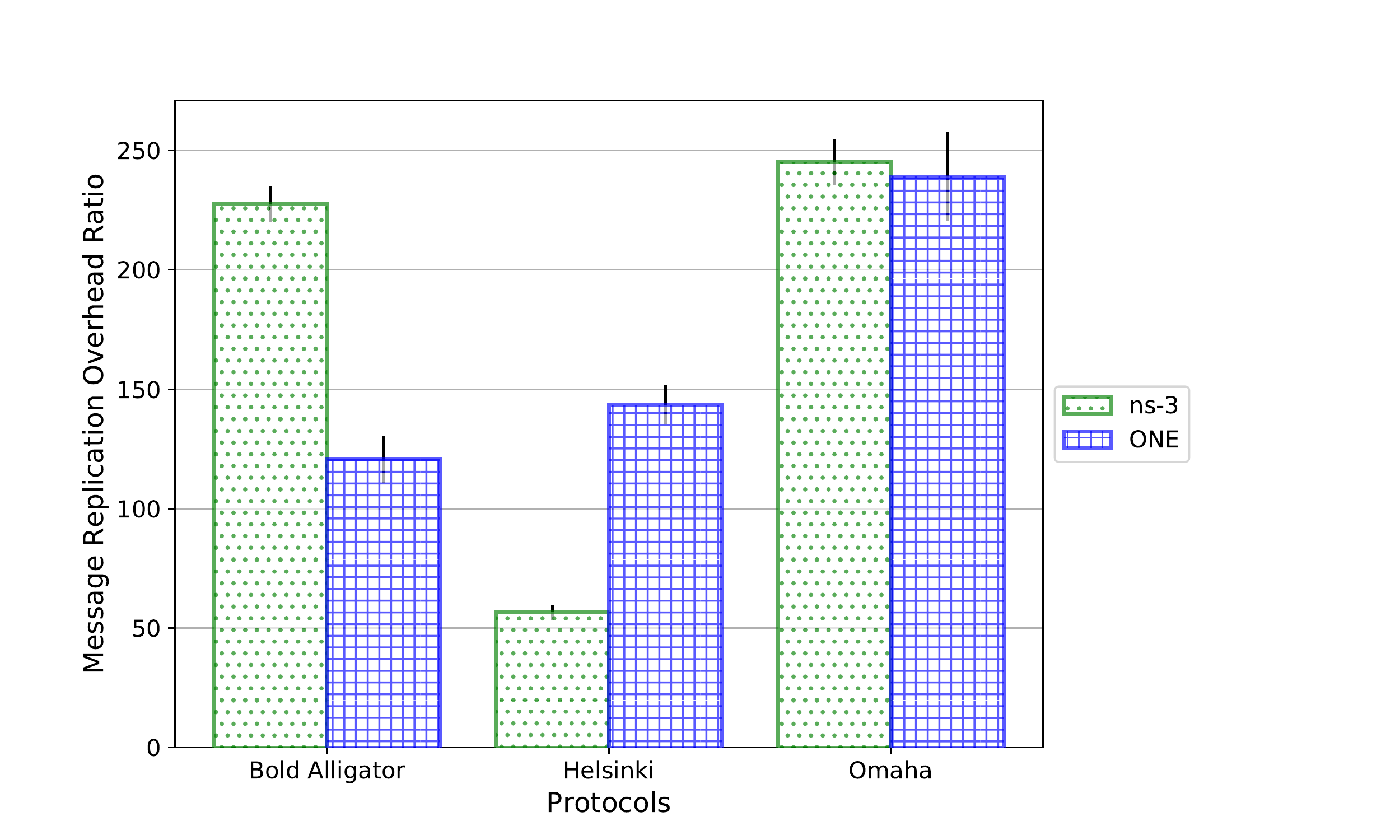}
	\label{fig:SimAggOvd}
	}
	\vspace{-12pt}
	\caption{Scenario aggregate performance}
\end{figure*}

\subsection{Aggregate Scenario Performance}
This section compares the overall performance of \ac{ns-3} and the \ac{ONE} using \ac{MDR}, average latency, average hop count, and message replication overhead.  
The bar graphs are an average of all buffer sizes, transmission speeds, and protocol data points for each scenario.  Each bar graph uses a 95\% confidence interval.  

Figure~\ref{fig:SimAggMDR} compares aggregate protocol and scenario \ac{MDR} between the \ac{ONE} and \ac{ns-3}.  Within Bold Alligator, \ac{ns-3} delivers 61\% fewer messages than the \ac{ONE}, and 24\% fewer messages in Omaha.  \ac{ns-3} delivers 17\% fewer messages in Helsinki.  Across all scenarios and protocols, \ac{ns-3} delivers 31\% fewer messages than the \ac{ONE}. 

Figure~\ref{fig:SimAggLat} compares aggregate average latency for protocols and scenarios between simulators.  Epidemic's average latency is 21\% higher in \ac{ns-3} than the \ac{ONE}.  
In relation to network overhead in Figure~\ref{fig:SimAggOvd}, protocols with higher message replication overhead tend to return a smaller difference between simulators.  Helsinki's average latency in \ac{ns-3} is 202\% higher than the \ac{ONE}.  Bold Alligator's average latency is 160\% higher in \ac{ns-3} and Omaha's average latency is 53\% higher in \ac{ns-3}.  Across all scenarios and protocols, \ac{ns-3}'s average latency is 119\% higher than the \ac{ONE}.

Unlike \ac{MDR} and average latency, message replication overhead in Figure~\ref{fig:SimAggOvd} displays different trends between military and urban scenarios.  \ac{ns-3} returns lower message replication overhead than the \ac{ONE} by 61\%.  However, \ac{ns-3}'s message replication overhead is 88\% higher in Bold Alligator and 2.5\% higher in Omaha.  We note that the \ac{ns-3} protocols increase message replication overhead due to messages circling within clusters of nodes.  The \ac{ONE} protocols, except for Epidemic, do not have this behavior because nodes check whether a message traversed the other node.  If the message already traversed the other nodes, then the node does not transmit the message.  Helsinki does not form node clusters, so \ac{ns-3}'s Helsinki does not return the higher message replication overhead.  

Messages circling within clusters does occur with the \ac{ONE}'s Epidemic, so the \ac{ns-3}'s message replication overhead is 65\% lower than the \ac{ONE}.  Across all scenarios and protocols, \ac{ns-3}'s message replication overhead is 5\% higher than the \ac{ONE}.

In summary, \ac{ns-3} delivers fewer messages and experiences higher average latency than the \ac{ONE}.  Protocols that share more data to make routing decisions tend to deliver fewer messages in \ac{ns-3} than the \ac{ONE}.  Message replication overhead depends on node mobility due to implementation differences.  Scenarios that form clusters of nodes return higher message replication overhead in \ac{ns-3}.  Scenarios that do not form clusters of nodes return lower message replication overhead in \ac{ns-3}.

\section{Conclusions}
\ac{ns-3} and the \ac{ONE} employ different levels of abstraction to simulate network protocols.  The \ac{ONE} focuses on simulating the behavior of opportunistic routing protocols, so the \ac{ONE} abstracts everything below the routing layer.  In contrast, \ac{ns-3} simulates the entire network stack.  The \ac{ns-3} routing protocols require packets for messages, node discovery, and sharing routing information between nodes.  The \ac{ONE} does not include link layer overhead, packet header overhead, or control packet overhead.

The \ac{ONE} sends \ac{DTN} data as a single object called a message, and the \ac{ONE} shares routing information by directly accessing communicating nodes' memory data structures.  Our \ac{ns-3} \ac{DTN} protocols assume an \ac{IP} convergence layer adapter by defining groups of packets that compose a \ac{DTN} message. Control packets share routing information between nodes in \ac{ns-3}. When comparing the effective throughput of messages transmitted between two connected nodes, the addition of packet headers and link layer overhead reduces effective throughput by 40\% to 70\% relative to the \ac{ONE}'s radio bandwidth.  Depending on the scenario and protocol, packet headers added during message segmentation in \ac{ns-3} make up 2\% to 5.5\% of all transmitted data.  Depending on the routing protocol and scenario, \ac{ns-3}'s control packets can consume a significant portion of transmitted data.  Control packets make up 0.1\% to 33\% of all transmitted data in \ac{ns-3}.  Protocols that share less information for routing decisions transmit fewer/smaller control packets.  While control packets contribute to network overhead, message replication represents a larger fraction of network overhead.  For protocols that use the same message limit algorithm, \ac{ns-3} protocols with more control packets consume more power.

When comparing the \ac{ONE} and \ac{ns-3} \ac{DTN} protocols, the \ac{ns-3} protocols returned 31\% lower \ac{MDR} and 119\% higher average latency aggregated across all protocols and scenarios.  Message replication overhead varies between scenarios.  In Helsinki, the \ac{ns-3} protocols return lower message replication overhead than the \ac{ONE}.  

Both simulators demonstrate sensitivity to buffer size.  Larger message buffers return higher \ac{MDR}.  In Helsinki, larger message buffers reduce average latency.  Bold Alligator and Omaha show that larger message buffers increase average latency.  \ac{ns-3} shows greater sensitivity to transmission speed than the \ac{ONE}.  In \ac{ns-3}, higher transmission speeds return higher \ac{MDR}, lower average latency, and increase message replication overhead.  Higher transmission speeds in the \ac{ONE} return small changes in \ac{MDR}, slightly reduces average latency, and increases message replication overhead.  \ac{ns-3}'s increased sensitivity to transmission speed is due to control packets, packet headers, and link layer overhead.

Based on our findings, our future \ac{DTN} protocol development will continue in \ac{ns-3} instead of the \ac{ONE}.  While the \ac{ONE} permits faster development of new protocols, and \ac{ns-3}'s simulations take 25 to 50 times longer than the \ac{ONE}, and \ac{ns-3} requires a separate program to analyze the large trace files; the \ac{ONE}'s abstraction may not reflect actual protocol performance.  The sharing of routing information via shared data structures subsidizes performance of protocols with very large control packets.
Considering control packets transmission impacts protocol performance analysis and link layer overhead significantly, and also significantly reduces message replication, as shown in our \ac{ns-3} results.

\begin{acks}
  The authors would like to thank Dr. Robert Beverly for his feedback and contribution
  to improving the rigor of this work.  We also thank the WNS3 reviewers for their constructive analysis that strengthened this contribution.

  The work was funded in part by the United States Marine Corps and United States Navy.
  Views and conclusions are those of the authors and should not be interpreted as representing the official policies or position of the U.S. government.
\end{acks}

\bibliographystyle{ACM-Reference-Format}
\bibliography{wns3.bib,../bib/jprohrer.bib,../bib/master.bib,../bib/rfc.bib}

\end{document}